\documentclass[twocolumn,showpacs, floatfix]{revtex4}
\usepackage{amsmath}
\usepackage{graphicx}
\usepackage{epsfig}
\usepackage{color}
\usepackage{soul}

\usepackage{txfonts}

\begin{document}
\title{Impact of rough potentials in rocked ratchet performance} 

\author{S. Camargo$^1$}
\email{sabrina.mga@gmail.com}
\author{C. Anteneodo$^{2,3}$}
\email{celia.fis@puc-rio.br}

\affiliation{
$^1$ EMAp, Funda\c{c}\~ao Get\'ulio Vargas, Rio de Janeiro, Brazil \\
$^2$ Department of Physics, PUC-Rio, Rio de Janeiro, Brazil  \\
$^3$ National Institute of Science and Technology for Complex Systems, Rio de Janeiro, Brazil\\
}

\begin{abstract}

We consider thermal ratchets modeled by overdamped Brownian motion in a spatially periodic
potential with a tilting process,  both unbiased on average.  
We investigate the impact  of the introduction of roughness in the potential profile,   
over the flux and efficiency of the ratchet. 
Both amplitude and wavelength that characterize roughness are varied.
We show that depending on the ratchet parameters, rugosity  can either spoil or   
enhance the ratchet performance.

\end{abstract}


\maketitle

\section{Introduction}

Some mechanisms working in the Brownian domain are not as intuitive as they may appear. 
One of them is ratcheting, where periodic forces which are null on average can still produce 
directed motion \cite{Reimann200257}.  
The study of this effect had initially biophysical motivations, 
like the operation of molecular motors
\cite{sozanski2015small,PhysRevLett.71.1477}, 
and more recently mainly technological ones, such as in the 
microfabrication of devices that can be used to  
separate or rectify the motion of microparticles
\cite{kettner2000drift,bader1999dna}. 
These implementations include, amongst other ones,  
silicon  membranes \cite{matthias2003asymmetric}, 
optical lattices \cite{lee2005flux,lee2005one,leon2017noise}, 
quantum dot arrays \cite{linke1999experimental} 
and vortex rectifiers in superconducting films \cite{cerbu2013vortex,villegas2003superconducting}.

There are mainly three types of ratchets: i) pulsating ratchets - the
potential is switched on and off or there is a traveling potential, changing
the barrier height; ii) tilting ratchets - with the addition of fluctuating
forces or a rocking periodic force, and iii) temperature ratchets \cite{Reimann200257}.
In our study, we choose a ratchet of the rocked type.
Nonequilibrium fluctuations are introduced as an additive tilting force, 
including the periodic rocking plus noise, all unbiased on average. 
Following the definitions of symmetry and supersymmetry~\cite{Reimann200257}, 
the net flux vanishes for any combination of potential and tilting which are both 
symmetric or both supersymmetric. Otherwise a net flux arises.
 
While the effect of spatiotemporal asymmetries, non-equilibrium fluctuations and other features   
have been extensively studied~\cite{Reimann200257, PhysRevLett.71.1477,Bartussek, Chialvo199526, PhysRevE.65.051103,
PhysRevE.70.021102, PhysRevLett.84.258,celia},  
the role of roughness in the periodic potential has been less explored. 
However, spatial inhomogeneities or impurities can yield 
deviations from a smooth ratchet profile, 
with important implications in transition rates.
In fact, the roughness of the potential is crucial in biophysical contexts such as in 
protein folding~\cite{frauenfelder1991energy,frauenfelder1999biological}, 
where the potential surface can exhibit a rich structure of maxima and minima, hierarchical or not.
The roughness in energy landscapes has been 
experimentally measured in proteins
\cite{volk2015roughness,milanesi2012measurement,nevo2005direct,kapon2008protein} but 
can also be microfabricated, for instance, holographic optical tweezers, used in optical ratchets, 
can generate complicated potential energy landscapes for Brownian particles
\cite{lee2007brownian}.

Within this scenario, Zwanzig~\cite{zwanzig1988diffusion} studied diffusion in a rough potential, 
pointing to the reduction of the effective diffusion coefficient when compared to a smooth surface.
Later, Marchesoni \cite{marchesoni1997} explored disorder
in a ratchet potential, due to impurities or randomness, reporting quenching of 
the effectiveness of thermal ratchets. 
More recently, Mondal et al.~\cite{mondal} showed that roughness hinders current significantly. 
In our work, we investigate the effects of perturbations of short wavelength 
superimposed on the periodic potential. 
Varying  the amplitude and wavelength of these perturbations,  
we monitor the net directed current, as well as the efficiency, 
for  a wide range of intensities of the time varying forces. 
We show that perturbations do not always spoil but, 
depending on the  ratchet parameters,  can enhance the performance.

The paper is organized as follows. 
We define the details of the ratchets and rugous perturbations of the potential in Section  \ref{sec:model}, 
and the methods in  Section  \ref{sec:methods}.
Results  for the effects of  sinusoidal perturbations, over   
symmetric and asymmetric forms of the spatial periodic potential, in the adiabatic limit,
are presented in Sections~\ref{sec:sUaF}-\ref{sec:aUaF} and also in the appendices. 
Other types of perturbations are analyzed in Section~\ref{sec:other}. 
The non-adiabatic regime is considered in Section~\ref{sec:nonad}. 
Concluding remarks are presented in Sec.~\ref{sec:conclusions}

\section{The system}
\label{sec:model}
We consider the overdamped regime for a particle of unit mass  obeying the equation of motion   
\begin{equation}
\dot{x}=-U'(x)+F(t)+\zeta(t)  \,,
\label{eq:motion2}
\end{equation} 
where $U(x)$ is a spatially periodic potential, $F(t)$ a time periodic driving,  
and  $\zeta(t)$  is a fluctuating force with zero mean, $\langle\zeta(t)\rangle=0$, and 
delta correlated, $\langle\zeta(t)\zeta(t')\rangle=2D\delta(t-t')$. 

The reference, or unperturbed, spatially periodic potential  is  
given by the saw-tooth shape~
\begin{equation}
U_0(x)=\left\{
\begin{array}{lr}
 \dfrac{x}{\ell}, & 0\le x<\ell,\\[0.5cm]
\dfrac{\lambda-x}{\lambda -\ell}, & \ell<x<\lambda,\\
\end{array}
\right.
\label{eq:potential0}
\end{equation}
where $\lambda$ is the spatial period, and $\ell$ controls the 
asymmetry of the potential, which is symmetric for $\ell=\lambda/2$. 
 We will set $\lambda=1$.

The periodic tilting protocol  is  
\begin{equation}
F(t)=\left\{
\begin{array}{lr}
 \dfrac{A}{\alpha}, & 0\le t<\dfrac{\alpha}{1+\alpha}\tau,\\[0.5cm]
-A, & \dfrac{\alpha}{1+\alpha}\tau <t\le\tau,\\
\end{array}
\right.
\label{eq:force}
\end{equation}
where $\tau$ is the time period, and the parameter $\alpha$ regulates the time symmetry of the force, 
which is symmetric when $\alpha=1$.

\begin{figure}[h!]
\centering
\includegraphics[scale=0.45]{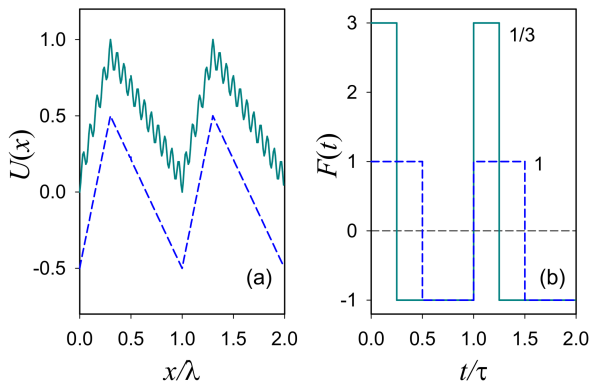}
\caption{(a) Potential given by  Eqs.~(\ref{eq:potential0}) 
and (\ref{eq:potential}), with $\ell=0.3$, for $\varepsilon=0.0$ (dashed),  
$\varepsilon=0.2$, $K=15$ and $N=1+\varepsilon$ (solid line).  
(b) Forcing given by  Eq.~(\ref{eq:force}), with $A=1.0$, for   
$\alpha=1$ (dashed)  and $\alpha=1/3$ (solid line).   
}
\label{fig:forms}
\end{figure}

These simple forms, illustrated in Fig.~\ref{fig:forms}, were chosen  because they 
allow easy symmetry control. 
Notice that the forces $-U_0'(x)$ and  $F(t)$ are unbiased on average, 
over one spatial and time period, respectively.

We will consider both regular and irregular perturbations  $u(x)$ of the reference potential $U_0$.   
We will start by analyzing a regular sinusoidal perturbation, 
 such that the total potential, depicted in Fig.~\ref{fig:forms}, is  
\begin{equation}
U(x)=  [U_0(x) - \varepsilon\cos(2\pi K x)/2 ]/N,
\label{eq:potential}
\end{equation}
where $\varepsilon<<1$ controls the amplitude of the perturbation and 
$K$ is an odd integer that controls its  wavenumber. 
We chose $K$ such that extremes of the perturbation coincide with extremes of the potential.
Moreover, $N$ is a normalization factor. 
In Section ~\ref{sec:results}, we will set $N= 1+\varepsilon$.   
With this choice, the largest barrier height $\Delta U= U(\ell)-U(\lambda)$ is kept unchanged.  
This implies a deformation of the  potential $U_0$ besides the addition of roughness.
For comparison, in the Appendix~\ref{appendix}, we will show the corresponding 
main results setting $N=1$, which means a purely additive perturbation of $U_0$.  
 
In order to specify the parameter space that will be investigated, 
 the values of $\ell$ and $\alpha$, that control the spatiotemporal symmetries, 
will be kept fixed for three different scenarios that produce net current in the unperturbed case 
(spatial asymmetry, temporal asymmetry, or both).  
The spatial period is fixed ($\lambda=1$). 
The parameter $\tau$, that controls the time periodicity,  will be varied only in the 
non-adiabatic case addressed in Section~\ref{sec:nonad}. 
Therefore, the parameter space has actually dimension four.  
Two of the parameters ($A$ and $D$) are associated to the unperturbed ratchet (tilting and noise amplitudes), 
while the other two ($\varepsilon$ and $K$) are associated to the rugosity (its amplitude and wavelength).  

Other kinds of perturbations (Weierstrass  and two-scale functions) will be also considered, as described  
in Sec.~\ref{sec:other}.

\section{Methods}
\label{sec:methods}

In the adiabatic limit, i.e., for  sufficiently large $\tau$  compared 
to the relaxation times of the system, 
the steady current can be obtained through \cite{Reimann200257}
\begin{equation}
 J  = \dfrac{1}{\tau}\int_{0}^{\tau} \langle \dot{x}\rangle \, dt = 
  \frac{\alpha J(A/\alpha) +   J(-A)}{\alpha+1}  \,,
	\label{Jad}
\end{equation}
where the current, at a constant driving  $F$,  is given by 
\begin{equation}
J(F) = \frac{D ( 1-{\rm e}^{-F/D}  )}{ \int_0^1  dy \int_0^1 dx \, {\rm e}^{- \Phi(y) + \Phi(x+y) } }  \,,
\label{JF}
\end{equation}
with $\Phi(x)= [U(x)-xF]/D$, after setting $\lambda=1$.

Another important quantity is the efficiency, that can be measured
as~\cite{efficiency} 
\begin{equation}
\eta = \frac{ {J}^2}{  E  } \,,
\label{eta0}
\end{equation}\
where the average energy is given by
\begin{equation}
  E =   \frac{A[   J(A/\alpha) - J(-A) ] }{\alpha+1}\,.
\label{barE}
\end{equation}

In the adiabatic limit, we obtained the current and efficiency through Eqs.~(\ref{Jad})
and (\ref{eta0}), respectively. 

For the non-adiabatic regime, Eq.~(\ref{eq:motion2}) was integrated by means of 
a stochastic fourth-order Runge-Kutta scheme \cite{hansen2006efficient}, using 
suitable integration step ($dt<10^{-3}$, remarking that smaller steps are required 
for shortest scale structures). 
Then, we computed the  average current  
\begin{equation}  \label{simulationsJ}
J=\langle\dot{x}\rangle=\langle-U'(x)+F(t)\rangle,
\end{equation} 
  over 20 cycles of the tilting protocol and over 100 trajectories, after a transient has elapsed.  
We also verified the agreement between the results from these simulations and 
Eqs.~(\ref{Jad})-(\ref{barE}), when using $\tau\simeq 50$, for which the adiabatic limit 
is nearly attained.

\section{Results: adiabatic limit}
\label{sec:results}

In this section, we will show results in the adiabatic regime, 
obtained from  Eqs.~(\ref{Jad})-(\ref{barE}), for different symmetries of the potential and tilting process  
that yield a net current.  
In all cases the potential is normalized using $N=1+\varepsilon$.

\subsection{Symmetric $U$, asymmetric $F$}
\label{sec:sUaF}

\begin{figure}[b!]
\centering
\includegraphics[scale=0.45]{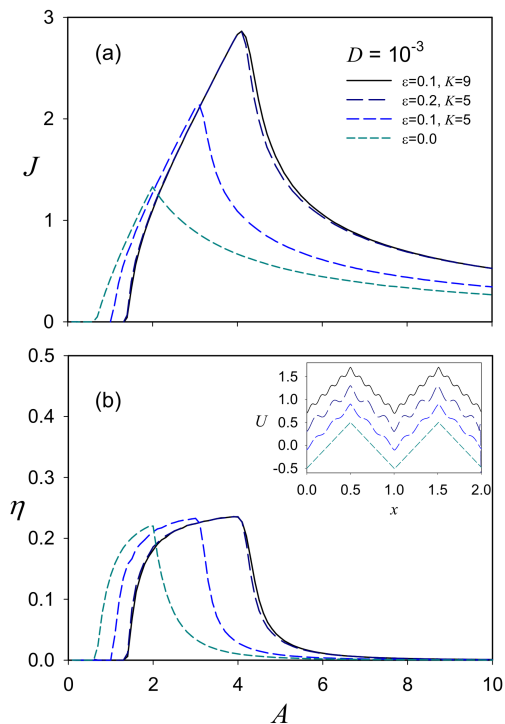}  \\[-3mm]
\caption{Symmetric $U$ (i.e., $\ell=0.5$) and  asymmetric $F$ (with $\alpha=1/3$). 
Current $J$ and efficiency $\eta$ versus force amplitude $A$, 
for the potential shapes given by Eq.~(\ref{eq:potential}) and drawn in the inset.     
$D=10^{-3}$. 
}
\label{fig:JvsA}
\end{figure}

Fig.~\ref{fig:JvsA} exhibits the current, $J$, and the efficiency, $\eta$, 
for asymmetric tilting ($\alpha=1/3$), and symmetric potentials ($\ell=0.5$) 
with different amplitudes and wavelengths of the perturbation, at low noise level.   
There is a value of $A$, the intersection point between the couple of curves with and 
without rugosity, below (above) which the rugosity spoils (enhances) the current. 
The same happens for the efficiency, although its critical value can be slightly different. 
As a matter of fact, the behavior at low $D$  can be understood 
by recalling the behavior in the deterministic limit ($D=0$): 
for very small $A$, the particle confined in a well cannot jump the 
barriers of the effective potential $U_{eff}=U(x)-xF(t)$, neither when the tilt 
is to the right nor to the left, then the  flux is always zero. 
From a critical value, let us call it $A_1$,  the particle can start to move 
in the positive direction (assuming $\alpha<1$ as in the case of Fig. \ref{fig:JvsA}), 
because the slope of the effective potential becomes negative for any $x$ 
when the potential is tilted to the right ($F>0$), 
while still it cannot move to the left. From that point, the current increases with $A$, 
until attaining a second critical value $A_2$, where the slope of $U_{eff}$ 
becomes positive for any $x$ when the tilt is to the left ($F<0$). 
Since the particle can move back and forth along one period,  
the net current starts to decrease. 
It continues decreasing monotonically 
with $A$, towards zero, since the potential structure becomes irrelevant in the large $A$ limit.

\begin{figure}[b!]
\centering
\includegraphics[scale=0.45]{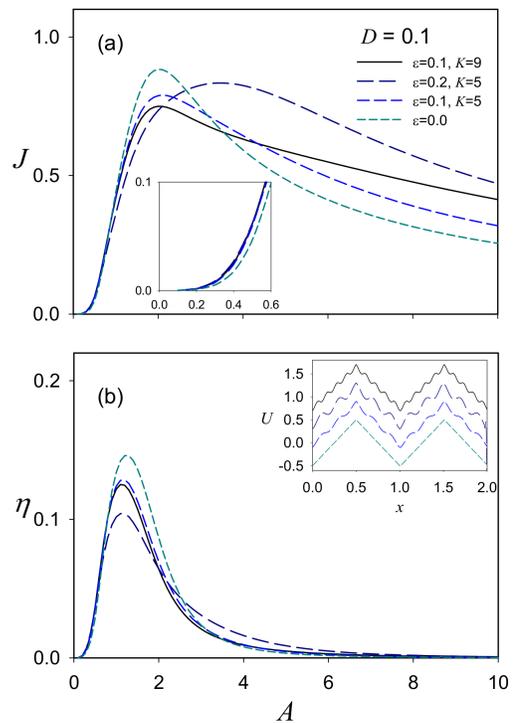}  \\[-3mm]
\caption{Symmetric $U$, asymmetric $F$. Current $J$ and efficiency $\eta$ versus force amplitude $A$, 
with the same parameters used in Fig.~\ref{fig:JvsA}, except that $D=0.1$. 
}
\label{fig:JvsA_RRR}
\end{figure}

When roughness is added, the presence of small wells traps the particle, 
leading to a larger critical value $A_1$ for current upraise in the positive direction. 
Similarly, the critical value $A_2$, for the upraise of current in the opposite direction, 
also increases. As a consequence, a larger maximal value is attained, because the particle 
moves only forth, while it moves back and forth in the unperturbed potential.

Also notice in Fig.~\ref{fig:JvsA}  that the cases $(K,\varepsilon)=(5,0.2)$ and 
$(9,0.1)$ present a quite similar behavior. This is because the maximal
slope of the potential, which compared to $xF$ determines the critical values in the small $D$ limit, 
is close in both cases.

However, when $D$ departs from the deterministic limit,  a  less trivial picture emerges  
(see Fig.~\ref{fig:JvsA_RRR}). 
For instance, the position and value of the largest current 
do not increase monotonically with the perturbation amplitude 
and wavenumber, as it happens for small $D$.  
Moreover,  the rugosity enhances  current and efficiency also at small $A$. 
This effect is not present in the deterministic limit, indicating the 
interplay between noise and roughness.

\begin{figure}[t!]
\centering
\includegraphics[scale=1.4]{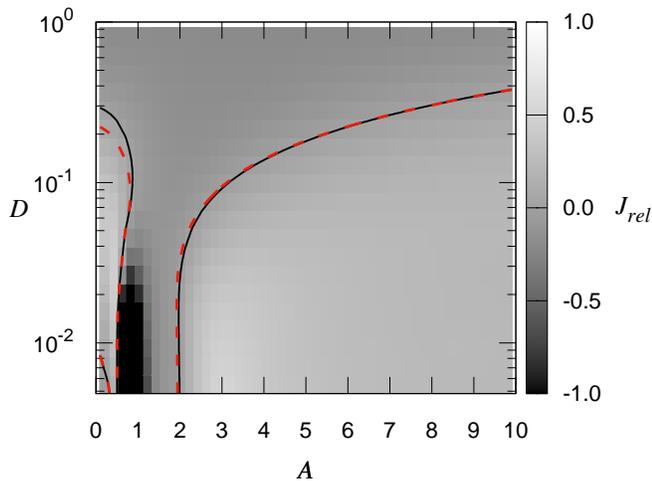}
\caption{Symmetric $U$ (i.e., $\ell=0.5$) and asymmetric $F$ (with $\alpha=1/3$). 
Gray-scale map of relative current difference $J_{rel}=(J_r-J_0)/J_0$ (comparing rugous and 
unperturbed cases) in the plane $A-D$. 
The perturbation is sinusoidal with $K=5$, $\varepsilon=0.1$. 
The solid  curve highlights the zero values of the relative current difference. 
The dashed line indicates zero values of the relative efficiency difference $\eta_{rel}$.
}
\label{fig:JvsAD}
\end{figure}

In order to provide a full portrait of the rugosity effects, 
we depict, in Fig.~\ref{fig:JvsAD}, a gray-scale map  in the plane of parameters $A-D$. 
We considered as representative example the perturbation with $K=5$ and $\varepsilon=0.1$.
For the sake of comparison between perturbed and unperturbed cases, 
instead of plotting  absolute quantities,  
we define the relative differences 
$J_{rel} = (J_r-J_0)/J_0$ and $\eta_{rel} = (\eta_r-\eta_0)/\eta_0$, 
where the subindices $r$ and $0$ indicate rugous and unperturbed profiles.

The roughness spoils the performance in the central (darker)  domain, 
within a limited interval of $A$ for small $D$ but for any $A$ when $D$ is large enough.
This regime was previously detected in Ref.~~\cite{mondal}, although for a different kind of
perturbation.
Differently, for small and moderate $D$, there are two  regions where the rugosity enhances the performance. 
Besides the region of large $A$,  predicted in the deterministic limit $D=0$, 
there is also a region of very small $A$ ($A \lesssim A_1$) where the performance 
is improved by the rugosity. 
A closer look on these effects is provided in the Appendix~\ref{appendix2}.

\subsection{Asymmetric $U$, symmetric $F$}
\label{sec:aUsF}

In this section, we analyze the impact of the rugosity when the 
potential $U_0$ is asymmetric while $F$ is symmetric.  
The asymmetry of the potential is controlled by parameter $\ell$. 
We will set $\ell=0.7$ (hence $K=5,15,\ldots$). 
At the same time, we set $\alpha=1$  to have a symmetric protocol $F(t)$.

\begin{figure}[h!]
\centering
 \includegraphics[scale=1.4]{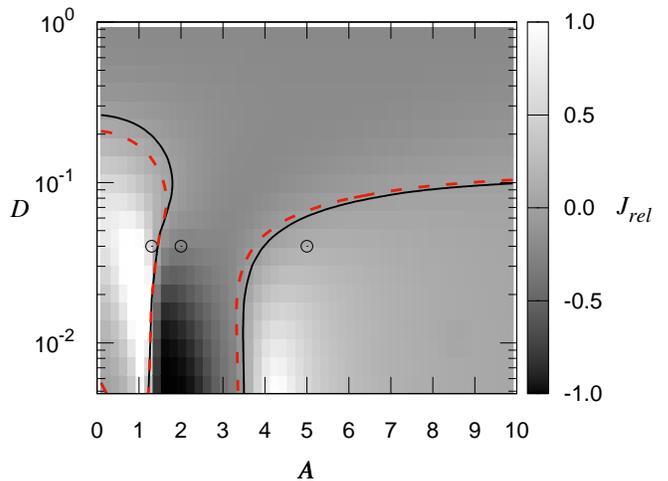}
\caption{
Asymmetric $U$ (with $\ell=0.7$) and  symmetric $F$ (i.e., $\alpha=1$). 
Gray-scale map of relative current difference in the plane $A-D$. 
The perturbation is sinusoidal with $K=5$, $\varepsilon=0.1$. 
The solid  curve indicates zero values of the relative current difference. 
The dashed line indicates zero values of the relative efficiency difference. 
The circles highlight the parameter values for which trajectories are discussed in  Appendix~\ref{appendix2}.
}
\label{fig:JvsADasym07}
\end{figure}

The results for a perturbation with $K=5$ and $\varepsilon=0.1$ are  summarized in    
the gray-scale map of the relative current difference in Fig.~\ref{fig:JvsADasym07}. 
The behavior of the current, $J$, and the efficiency, $\eta$, as a function of $A$ 
presents similar features as those shown in   Sec.~\ref{sec:sUaF}. 
Also in this case roughness is able to increase the performance, in domains similar to 
those of Fig.~\ref{fig:JvsAD}.
Notice that net flux enhancement occurs  despite the partial loss of asymmetry 
of the potential due to the introduction of a symmetric perturbation and normalization.

\subsection{Asymmetric $U$ and $F$}
\label{sec:aUaF}

In this subsection, we analyze the impact of the rugosity when both the 
potential $U_0$ and $F$ are asymmetric. 
The asymmetry of $F$ will be given by $\alpha=1/3$ and the 
asymmetry of $U$ by choosing $\ell=0.1$ (hence $K=5,15,\ldots$ are allowed). 
With this choice,  the asymmetries have opposite tendencies, that is, promote net currents in opposite 
directions. Therefore, current inversions occur, for instance, as a function of $A$.

\begin{figure}[h!]
\centering
\includegraphics[scale=0.45]{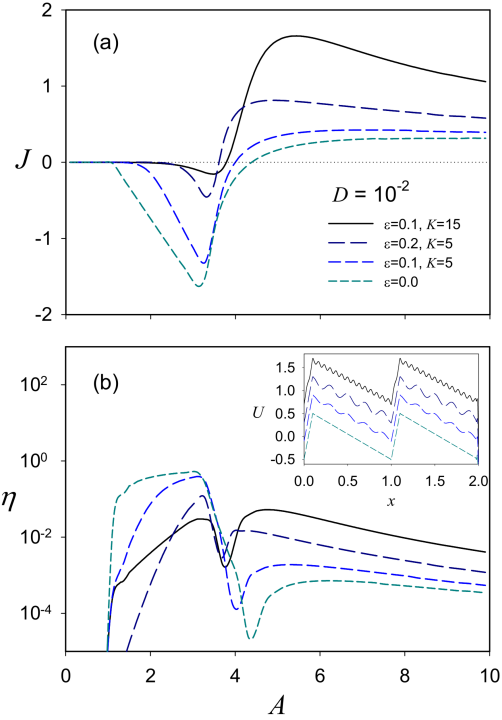}  \\[-3mm]
\caption{Asymmetric $U$ (with $\ell=0.1$) and asymmetric $F$ (with $\alpha=1/3$). 
Current $J$ and efficiency $\eta$ versus force amplitude $A$, for different potential shapes 
given by Eq.~(\ref{eq:potential}) and shown in the inset, at low noise amplitude $D$. 
}
\label{fig:JvsAasym2}
\end{figure} 

\begin{figure}[h!]
\centering
\includegraphics[scale=1.4]{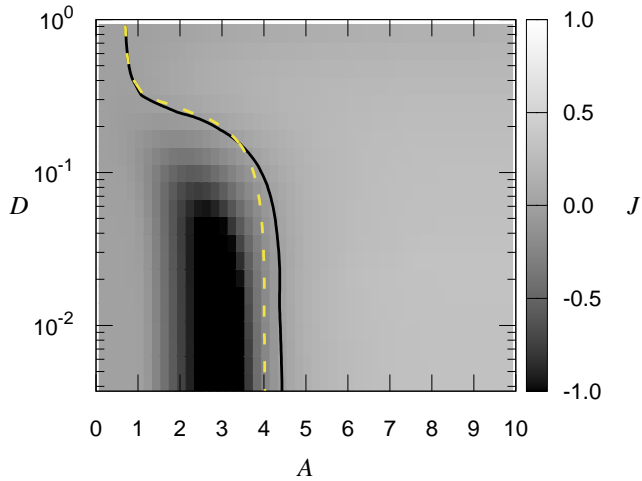}
\caption{Asymmetric $U$ (with $\ell=0.1$) and asymmetric $F$ (with $\alpha=1/3$).  
Gray-scale map of the current $J$  in the plane $A-D$, in absence of perturbation.  
The solid line indicates zero values. The dashed line indicates zero values when 
the perturbation is sinusoidal with $K=5$ and $\varepsilon=0.1$.
}
\label{fig:JvsADreversal}
\end{figure}

The effect of roughness parameters over current and efficiency is illustrated in Fig.~\ref{fig:JvsAasym2}.
In this case the introduction of roughness can produce current reversal, for fixed $A$, 
as observed in Fig.~\ref{fig:JvsAasym2}a, for instance around $A=4$. 
That is, roughness shifts the critical points at which net flux changes sign. 
Because of current inversion, we represent in Fig.~\ref{fig:JvsADreversal}, 
a gray-scale map in the plane $A-D$ of current $J$ itself 
(instead of the relative current difference shown in previous maps), 
highlighting the points of current reversal,  
for both the unperturbed and perturbed potentials. 
Notice that, for instance, at fixed $D<0.1$ and  $A\approx 4.2$ the removal of the rugosity produces current inversion.

When the potential is not normalized (i.e., $N=1$), 
there are also regions where roughness produces inversion of the current, but their frontiers change 
(see Appendix \ref{appendix}).


\subsection{Other perturbations}
\label{sec:other}

In order to check the robustness of the results, 
instead of the sinusoidal perturbation, we considered the following ones.  

i) Based on the hierarchical Weierstrass function  
truncated at order $n$, we define
\begin{equation} \label{u1}
U(x,n)= \sum_{j=0}^n a^j \cos(2\pi b^j x)/\biggl(2\sum_{j=0}^n a^j \biggr)  \,, 
\end{equation}
setting $b=3$ and $a=0.7$.

ii) We also considered a two-scale perturbation,  based in Refs.~\cite{mondal,kurths}, which is a
superposition of sine and cosine functions with different wavelengths, much smaller than $\lambda$, the 
spatial period of $U_0(x)$, namely, 

\begin{equation} \label{u2}
u(x)=-\varepsilon [\sin(85x)+ \cos(57x)] \,.
\end{equation}

\begin{figure}[h!]
\centering
\includegraphics[scale=0.45]{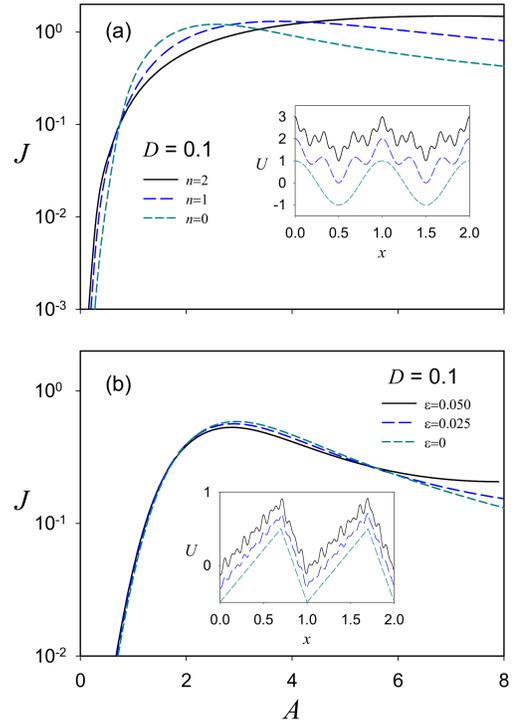}   
\caption{Other perturbations. 
Current $J$  versus force amplitude $A$,  for: 
(a) normalized potential defined in Eq.~(\ref{u1}), with $a=0.7$, $b=3$,  and different hierarchies $n$, 
plus asymmetric tilting, with $\alpha=1/3$, and 
(b) $U_0$ defined by Eq.~(\ref{eq:potential0}) with $\ell=0.7$ 
and perturbation given  by Eq.~(\ref{u2}), with values of $\varepsilon$ indicated on the figure,  plus 
 symmetric tilting (i.e.,$\alpha=1$). In both cases  $D=0.1$. 
}
\label{fig:irregular}
\end{figure}

The effects of these perturbations are illustrated in Fig.~\ref{fig:irregular} for $D=0.1$.  
In both cases, the current increases with increasing hierarchy $n$ or amplitude $\varepsilon$, 
except in the intermediate range of $A$ around the maximum, 
as signaled by the two intersections of each couple of curves obtained with and without rugosity.

\section{Non-adiabatic regime}
\label{sec:nonad}

\begin{figure}[h!]
\centering
\includegraphics[scale=1.1]{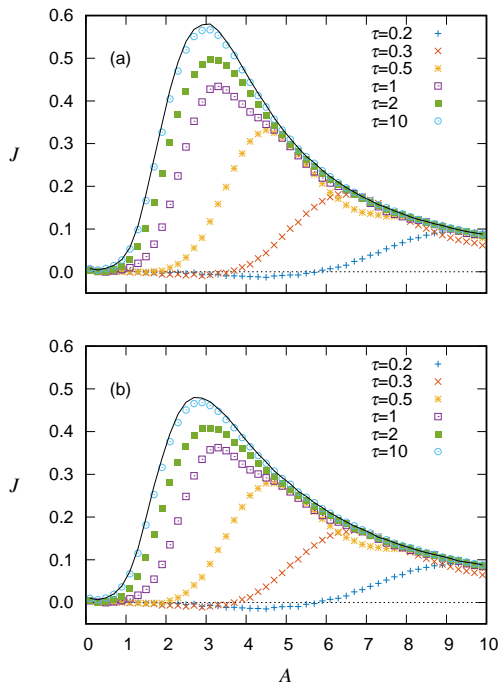}   
\caption{Non-adiabatic regime. 
Current $J$  versus force amplitude $A$,  for the 
asymmetric potential ($\ell=0.7$) and symmetric driving ($\alpha=1$). 
The perturbation is sinusoidal with $K=5$, with  
(a)  $\varepsilon=0$ (unperturbed), and 
(b) $\varepsilon=0.1$ (normalized, with $N=1+\varepsilon$).
Symbols correspond to numerical integration of Eq.~(\ref{simulationsJ}), 
the solid line to the theoretical values given by Eq.~(\ref{Jad}) valid in the adiabatic-limit.
In both cases, noise level is $D=0.1$.
}
\label{fig:non-ad}
\end{figure}

Finally, we also investigated the non-adiabatic regime. 
In this case we used numerical integration of the equation of motion, as described in Sec.~\ref{sec:methods}.
We focus on the case of Sec.~\ref{sec:aUsF}, where the potential is asymmetric ($\ell=0.7$) 
and the tilting force symmetric ($\alpha=1$), while the perturbation is given by 
$K=5$, $\varepsilon=0.1$. Additionally, we consider different values of $\tau$. 
In Figs.~\ref{fig:non-ad}a and ~\ref{fig:non-ad}b, 
we plot the current $J$ versus $A$,  using different values of $\tau$, 
for the unperturbed ($\varepsilon=0$)  and normalized (with $\varepsilon=0.1$) cases, respectively. 
For $\tau \gtrsim 10$,  the adiabatic limit is practically attained. 
For small $\tau$, below a critical value, current inversion emerges for a range of $A$ in the unperturbed case. 
The maximal value of the current is diminished and the negative current slightly augmented by the rugosity.
In the additive case, differences are still smaller.
Moreover, the discrepancies between the unperturbed and perturbed cases decrease as $\tau$ decreases.

\section{Conclusions}
\label{sec:conclusions}

We investigated how small perturbations that incorporate rugosity on $U(x)$ 
affect the current in the adiabatic limit 
and in non-adiabatic regimes. 
We considered two cases, one where the potential is scaled ($N=1+\varepsilon$), 
maintaining the largest barrier height, another where the perturbation is 
purely additive ($N=1$). Their effects were shown in Sec.~\ref{sec:results} and Appendix~\ref{appendix}, 
respectively. 
In both cases, there are regions in parameter space where perturbations 
improve the current and efficiency.  
In particular, but not only, for large $A$, as predicted in the deterministic limit. 
In the scaled case, there is a region for small $A$ and $D$ where rugosity enhances 
the net current. Such region is absent when the perturbation is purely additive, in which 
case the  region of large $D$ is favored. 
In Sec.~\ref{sec:sUaF} and Appendix~\ref{appendix2}, we provided  physical insights about the origin of the observed 
effects. 
The roughness can also alter current reversal when both the potential and 
tilting are asymmetrical and the ratchet parameters are kept fixed. 
In the non-adiabatic regime, there are also variations due to the rugosity, but they decay with 
decreasing $\tau$, that is moving away from the adiabatic limit.

Although we studied systematically these effects mainly for  sinusoidal perturbations of different 
amplitude and wavelength, we have also shown that the same tendencies occur for other perturbations, 
namely, 
a hierarchical Weierstrass function, and a two-scale perturbation, with separated scales much 
smaller than  the spatial period of the potential.

\begin{acknowledgments}
SC acknowledges FGV and CA acknowledges Faperj (Foundation for Research
Support, State of Rio de Janeiro) and CNPq (National Council for Scientific
and Technological Development).
\end{acknowledgments}

\appendix

\section{Non-normalized potentials}
\label{appendix}

\begin{figure}[h!]
\centering
\includegraphics[scale=1.4]{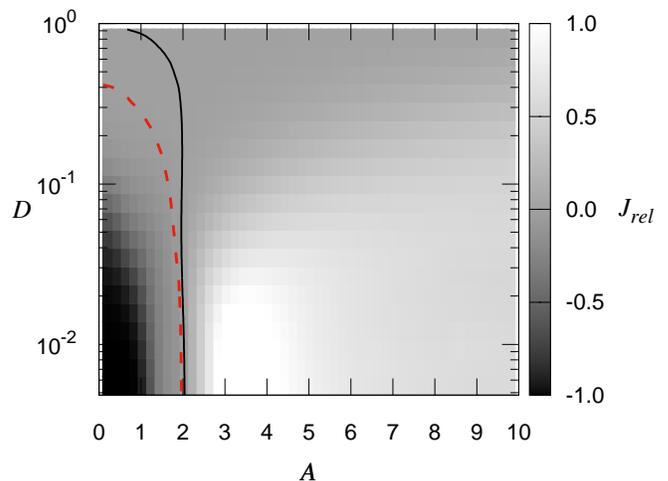}
\caption{Gray-scale map of relative current difference $J_{rel}=(J_r-J_0)/J_0$ in the plane $A-D$, 
for the same parameters used in Fig.~\ref{fig:JvsAD}, except that  $N=1$.
}
\label{fig:JvsADNN}
\end{figure}

\begin{figure}[h!]
\centering
 \includegraphics[scale=1.4]{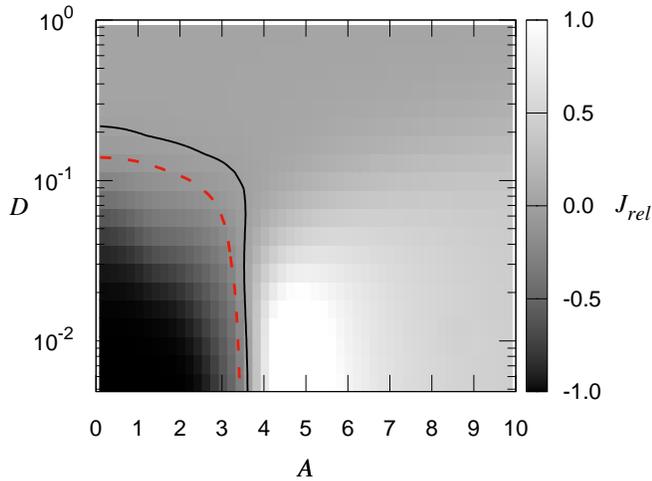}   
\caption{Gray-scale map of relative current $J_{rel}$ in the plane $A-D$, 
for the same parameters used in Fig.~\ref{fig:JvsADasym07}, except that $N=1$.
}
\label{fig:JvsADasym07NN}
\end{figure}

In this section, we summarize results for the purely additive perturbation, setting
$N=1$. 
Figs.~\ref{fig:JvsADNN} and \ref{fig:JvsADasym07NN} are the analogous of 
Figs.~\ref{fig:JvsAD} and \ref{fig:JvsADasym07}, obtained for $N=1+\varepsilon$.
%
%

In both cases, there are also regions in the plane $A-D$ where roughness is beneficial, 
but their frontiers are different.
In Figs.~\ref{fig:JvsADNN} and \ref{fig:JvsADasym07NN}, the region at low $A$ and small $D$ has disappeared, 
while the region of large $A$ is enlarged, standing for any $D$ and even smaller values of $A$ at large $D$. 
However, $J_{rel}$ is nearly null.

%
%
\begin{figure}[h!]
\centering
\includegraphics[scale=1.4]{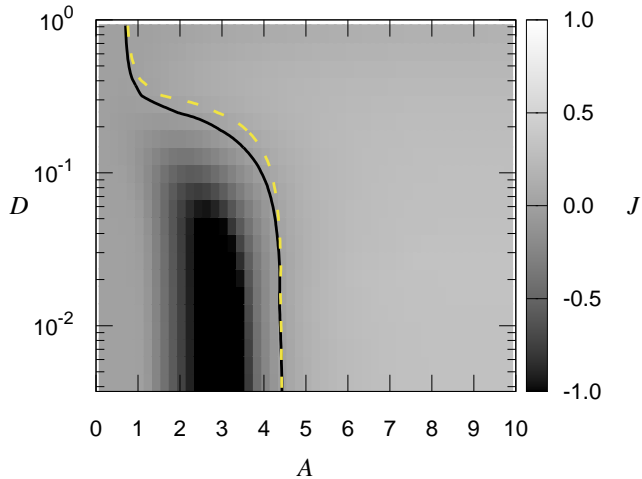}
\caption{Gray-scale map of the current $J$  in the plane $A-D$,   
like in Fig.~\ref{fig:JvsADreversal}, with $N=1$.
}
\label{fig:JvsADreversalNN}
\end{figure}

When both potential and tilting are asymmetric, the domain of current 
reversal in the presence of roughness has also changed 
(compare Figs.~\ref{fig:JvsADreversal} and ~\ref{fig:JvsADreversalNN}). 



\section{A closer look}
\label{appendix2}

\begin{figure}[h!]
\centering
\vspace*{5mm}
\includegraphics[scale=1.3]{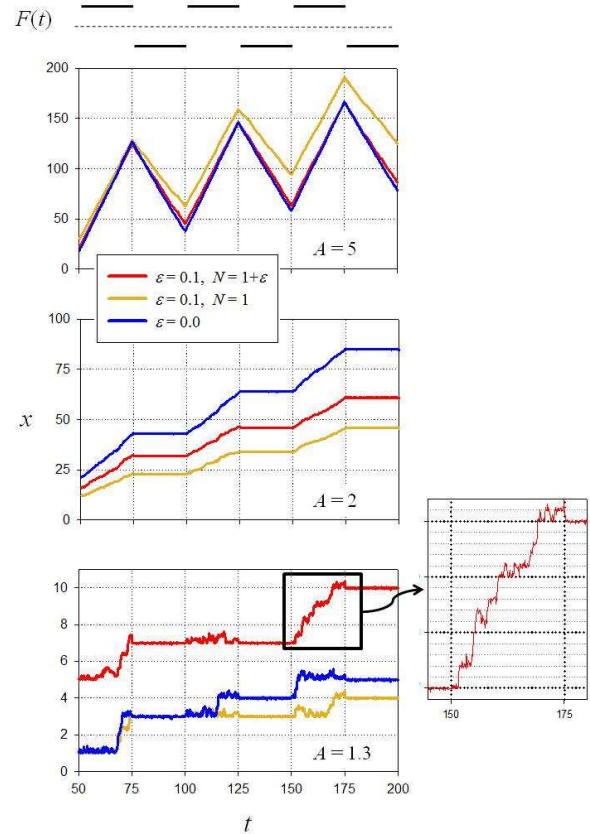}    
\caption{Trajectories for the  asymmetric potential ($\ell=0.7$) with symmetric drift ($\alpha=1$).  
The values of $A$  are indicated in the panels, and in all cases $D=0.04$  
(these cases, marked in Fig.~\ref{fig:JvsADasym07}, are illustrative of each region in the $A-D$ plane). 
We show three time periods of the movement under the  two perturbed and the unperturbed potentials, 
when the perturbation is sinusoidal with $K=5$ and $\varepsilon=0.1$.  
The tilting protocol $F(t)$ (with $\tau=50$) is also depicted at the top, for comparison. 
For $A=1.3$, a zoom is shown, for the trajectory in a normalized potential.
}
\label{fig:trajectories}
\end{figure}

In order to have a closer look on the differences observed, 
for instance between Figs. \ref{fig:JvsADasym07} and \ref{fig:JvsADasym07NN}, 
we plot in Fig.~\ref{fig:trajectories}, 
trajectories $x$ versus $t$ for the case of  an asymmetric potential ($\ell=0.7$) with  symmetric drift ($\alpha=1$), 
unperturbed and perturbed, with and without normalization.  
The perturbation is sinusoidal with $K=5$, $\varepsilon=0.1$.  
We consider representative points, in each region of Fig.~\ref{fig:JvsADasym07}, marked with a circle.

%

Let us define the back and forth currents, 
$J^-=-J(F<0)$, when the tilt is positive, and $J^+=J(F>0)$, when the tilt is negative.

i) For large $A$ (e.g., $A=5$ in the figure), the back  and forth currents are comparable. 
Both are slightly hindered by the rugosity, which introduces small barriers. 
But, due to the asymmetry of the potential with $\ell>0.5$ (the same holds for a tilting asymmetry with $a<1$), 
$J^-$ is more affected. This can be seen in the example of Fig.~\ref{fig:trajectories}, where 
discrepancies (between unperturbed and rugous cases) are augmented when the tilting is negative. 
The consequence is a larger net current in 
the presence of rugosity.
From this viewpoint, the effect is accentuated by increasing $\varepsilon$, because traps become deeper, and 
by increasing $K$, which makes trapping events more frequent.

ii) For the intermediate case $A=2$, since $A\gtrsim A_1$, the net current is dominated by $J^+$, 
that is $J^-$ is almost negligible when $F<0$ (associated to  the flat segments of $x(t)$). 
But  $J^+$  is spoiled by the small barriers, that produce segments of $x(t)$ with smaller slope under 
perturbations of the potential.  Consequently, in contrast to the previous case, the current decreases 
with rugosity.

iii) For small $A<A_1$  (e.g., $A=1.3$), and low noise level, the current is typically very small. 
Then, the current  to the left $J^-$ is negligible (flat segments of $x(t)$), 
the current to the right $J^+$ is also very small, but larger in the rugous case. 
This can be understood as follows. 
During the intervals of $F>0$, while the small valleys set an obstacle to the positive movement, 
they also block the movement backwards in larger extent, and the  particle can manage to pass to the next well 
jumping from  successive small wells (as can be seen in the zoom panel, where the trajectory stabilizes 
for short intervals in the positions of the small valleys). 
This can not occur in the smooth potential, or in the absence of noise. 
In the additive case, this subtle effect can occur, although more rarely, 
but, apparently, it is compensated by the largest potential difference.
%


\end{document}